\documentclass[sigconf]{acmart}

\AtBeginDocument{%
  \providecommand\BibTeX{{%
    \normalfont B\kern-0.5em{\scshape i\kern-0.25em b}\kern-0.8em\TeX}}}

\copyrightyear{2024}
\acmYear{2024}
\setcopyright{rightsretained}
\acmConference[GLSVLSI '24]{Great Lakes Symposium on VLSI 2024}{June 12--14, 2024}{Clearwater, FL, USA}
\acmBooktitle{Great Lakes Symposium on VLSI 2024 (GLSVLSI '24), June 12--14, 2024, Clearwater, FL, USA}
\acmDOI{10.1145/3649476.3658808}
\acmISBN{979-8-4007-0605-9/24/06}

\usepackage{algpseudocode}
\usepackage[linesnumbered,ruled]{algorithm2e}
\usepackage{makecell}
\usepackage{multirow}

\begin{document}

\title{Accelerating Boolean Constraint Propagation for Efficient SAT-Solving on FPGAs}


\author{Hari Govindasamy}
\affiliation{%
  \institution{Carleton University}
  \city{Ottawa}
  \country{Canada}}
\email{hari@sce.carleton.ca}

\author{Babak Esfandiari}
\affiliation{%
  \institution{Carleton University}
  \city{Ottawa}
  \country{Canada}
}
\email{babak@sce.carleton.ca}

\author{Paulo Garcia}
\affiliation{%
 \institution{Chulalongkorn University}
 \city{Bangkok}
 \country{Thailand}}
\email{paulo.g@chula.ac.th}


\renewcommand{\shortauthors}{Govindasamy, Esfandiari, and Garcia}

\begin{abstract}
  We present a hardware-accelerated SAT solver targeting processor/Field Programmable Gate Arrays (FPGA) SoCs. Our solution accelerates the most expensive subroutine of the Davis-Putnam-Logemann-Loveland (DPLL) algorithm, Boolean Constraint Propagation (BCP) through fine-grained FPGA parallelism. Unlike prior state-of-the-art solutions, our solver eliminates costly clause look-up operations by assigning clauses directly to clause processors on the FPGA and dividing large formulas into smaller partitions manageable by FPGA. Partitions are hot-swapped during runtime as required and the supported formula size is limited only by available external memory, not on-chip FPGA memory.
\par We evaluate our solver on a Xilinx Zynq platform with results showing quicker execution time across various formula sizes, subject to formula partitioning strategy. Compared to prior state-of-the-art, we achieve 1.7x and 1.1x speed up on BCP for 2 representative benchmarks and up to 6x total speedup over software-only implementation.
\end{abstract}





\maketitle

\section{Introduction}
The Boolean Satisfiability problem (SAT) is a fundamental problem in computer science, the first NP-Complete problem \cite{10.5555/574848}. SAT solvers have become the backbone of several engineering domains, as any NP-Complete problem can be encoded as instance of SAT \cite{10.5555/574848, cook2023complexity}. SAT solvers determine whether a given boolean formula is satisfiable by identifying an assignment to the formulas' free variables that evaluate the formula to true. The formula is unsatisfiable otherwise. Most SAT solvers target CNF-SAT, a subset of SAT that determines the satisfiability of formulas encoded in \emph{Conjunctive Normal Form} (CNF) . Formulas in CNF are conjunctions of clauses, where each clause is a disjunction of one or more literals (a variable or its negation).
\par With the advent of modern Systems-on-Chip (SoC) comprised of both hard embedded processors and configurable FPGA fabric offering myriad implementation opportunities \cite{stewart2019verifying}, deployed from the embedded to the high performance computing domain \cite{atitallah2017fpga}, accelerating SAT-solving through hardware is an attractive approach. We present a novel architecture for hardware-accelerated SAT-solving that outperforms state of the art solutions, released in open-source form for the Xilinx Zynq platform. Specifically, this article offers the following contributions:

\begin{itemize}
    \item We describe a methodology to map and runtime-manage clauses across a processor and connected FPGA, making efficient use of FPGA resources and avoiding recurring performance pitfalls.
    \item We describe the implementation of an open-source prototype system, deployed on a Xilinx Zynq chip, identifying how the hardware architecture effects the aforementioned strategy.
    \item We evaluate our design against the state of the art using two representative benchmarks, showing speed-ups of 1.7x and 1.1x, respective, and overall a 6x improvement over vanilla software execution.
\end{itemize} 

\par Section \ref{sec:back} describes necessary background knowledge on a particular SAT-solving algorithm required to understand the remainder of this paper. Section \ref{sec:art} presents an overview of historical solutions and state of the art, directly compared against in this paper. Section \ref{sec:arch} presents our contribution, evaluated in Section \ref{sec:results}, with concluding remarks and suggestions for future work described in Section \ref{sec:conclusions}.

\begin{figure}[t!]
\includegraphics[width=\columnwidth]{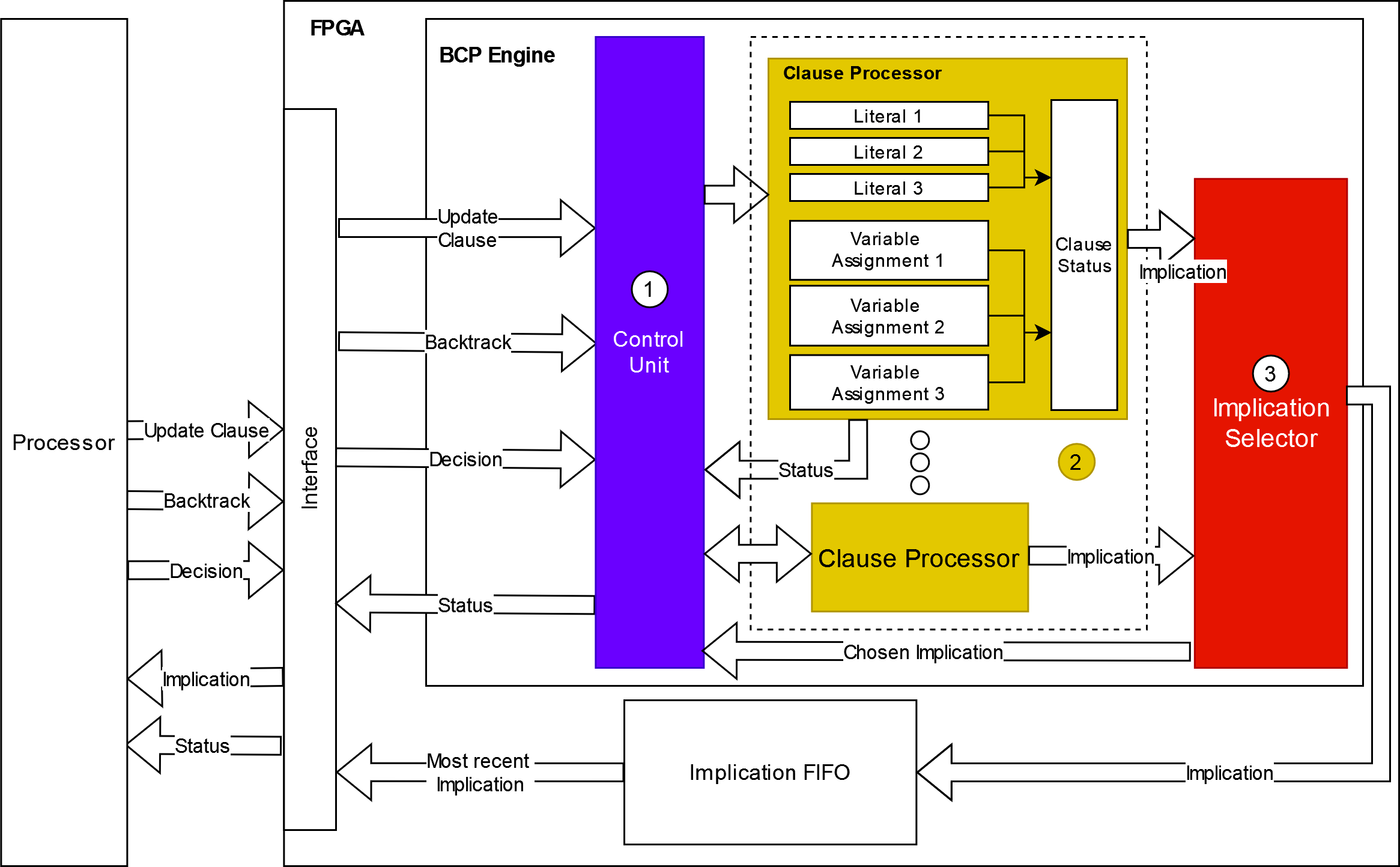}
\centering
\caption{Interface between processor and FPGA-based BCP Coprocessor. DPLL's BCP is accelerated through fine-grained parallelization across Clause Processors.}
\label{fig:processor_fpga_interface}
\end{figure}

\section{Background: DPLL and BCP}\label{sec:back}
\par SAT solvers are categorized into complete and incomplete solvers\cite{1288177}. Complete solvers evaluate every possible variable assignment, ending on the first satisfying assignment or after exhausting the search space. A formula is unsatisfiable if the complete solver concludes without finding a satisfying assignment.  Most incomplete solvers use Stochastic Local Search (SLS) to greedily search for a satisfying assignment in the formula's variable assignment search space\cite{SOHANGHPURWALA2017170}. While typically quicker than complete solvers, incomplete solvers do not guarantee results as they tend to get stuck in local maxima or skip satisfying assignments. Since they don't explore the solution space exhaustively they can never conclude that a formula is unsatisfiable. Davis-Putnam-Logemann-Loveland (DPLL) and DPLL-based algorithms are the most predominant complete solvers \cite{SOHANGHPURWALA2017170}. DPLL performs two primary operations: 1) \emph{decision} and 2) \emph{Boolean Constraint Propagation} (BCP). DPLL-based algorithms follow DPLL's core structure, and propose improved decision heuristics, learning and BCP mechanisms.  During decision, DPLL heuristically picks and assigns truth values to free variables. BCP subsequently propagates the effect of the decision using the \emph{unit implication rule}\cite{4555925}. The unit implication rule identifies unit clauses where all but one of its literals are false. Unit clauses can only be satisfied by assigning the variable to true if the literal is positive or to false on negative literals. The resulting assignment is known as an implication. BCP repeatedly applies this rule until all clauses are satisfied (formula is therefore satisfiable) or at least one clause evaluates false (conflict). On conflicts, DPLL backtracks by retracting and/or inverting assignments from earlier decisions.  BCP is expensive, accounting for 80-90\% of DPLL's CPU time, rendering it a prime candidate for hardware acceleration \cite{4555925,6691124}. BCP coprocessors accelerate DPLL by implementing specialized BCP processing engines on FPGA. These run alongside a General Purpose Processor (GPP) that performs the remaining DPLL operations: decision heuristics and backtracking. Using this architecture, the BCP coprocessor is first configured with the clauses, and then waits to evaluate decisions from the GPP. Any DPLL-based software solver can integrate with a BCP-coprocessor by replacing software BCP with the hardware accelerated BCP-coprocessor \cite{6691124,7372575}. FPGA-based BCP coprocessors are either instance-specific or application-specific. Instance-specific solver are built to solve a single SAT instance and designed by translating an input formula into its equivalent logical circuit. However, to solve new instances, the FPGA requires a complete rebuild (synthesis and FPGA programming may take up to several hours). Although these solvers can be significantly quicker than their software counterparts, their performance becomes notably slower when build times are included. For instance, Ivan et al's best result against the \textit{hole7} benchmark achieves a 6.66x speedup against MiniSAT\cite{6628279, 6601319}; however, when build times are included, compilation alone takes 50 seconds, whereas MiniSAT finishes in under 0.064 seconds \cite{8431977}. Application-specific solvers eliminate the need to rebuild the FPGA by instantiating general-purpose processing units capable of tackling any SAT instance (given that it fits in hardware). The BCP coprocessor is configured with the target problem by simply overwriting FPGA memory.

\section{State of the art}\label{sec:art}
Algorithmic techniques for efficient SAT solving have been extensively researched, and the literature contains several surveys that describe the history and state of the art of the problem (\cite{gong2017survey}, \cite{martins2012overview}). Techniques aimed at accelerating the execution of a particular SAT solving algorithm include software parallelization \cite{hamadi2010manysat}, deployment on specialized GPUs \cite{osama2021sat}, and even acceleration through machine-learning approaches \cite{wu2017improving}.
\par Our approach sits within FPGA-based acceleration, which began roughly 3 decades ago \cite{farrahi1994fpga}, with a few prominent results at the turn of the century (\cite{zhong2000fpga}, \cite{dandalis2002run}). However, it was not until significant advances in FPGA performance occurred in the last decade, and the rise of SoC platforms combining FPGA fabric with hard processors, that FPGA-based SAT acceleration matured. The most notable architectures were proposed by Davis et al \cite{4555925} and Thong et al \cite{6691124, 7372575}: both exploring the use of FPGA to implement BCP coprocessors, keeping the remainder of DPLL in software.
\par Davis et al calculate implications in parallel by using several inference engines (IE), each assigned a list of clauses (partitions) \cite{4555925}. For every decision/implication, the clause containing the assignment variable is first retrieved before calculating implications. Implications are forwarded to a conflict detector that ensures that two or more IEs have not implied opposing values for the same variable. Implications are then sent to the processor and queued up for propagation.
\par To keep clause retrieval time low, a variable only occurs once in each IEs partition (i.e clauses within the same IE share no common variables). This limits the number of clauses affected by a decision to one, thereby also limiting implications per IE to one, constraining the effected performance. While some strategies to increase this limit have been proposed \cite{davis2008designing}, they remain unexplored.
\par Thong et al. propose a concurrent BCP coprocessor comprising multiple sequential processing engines (PE) \cite{7372575}. Identifying that Davis et al.'s clause lookup is slower than direct access \cite{6691124}, they develop a clause storage and encoding scheme that efficiently links clauses with shared variables. The processor sends decisions to the FPGA and starts BCP execution at a single PE. Using the linked list, the PE traverses every clause containing the decision variable and calculates implications, which are then added to a local queue and propagated. The running PE triggers BCP execution in another PE when it arrives at a link to a clause that is located elsewhere. The coprocessor supports multithreaded software execution, hiding communication and software latency by keeping the coprocessor busy while software threads make decisions when possible.
\par Davis et al. and Thong et al. have laid a strong foundation in developing application-specific FPGA-based BCP coprocessors; we extend their work and propose a solution that processes clauses in parallel without the need for clause lookup.

\section{The SAT solver Architecture}\label{sec:arch}
We present a BCP coprocessor that works alongside vanilla DPLL (and should, in theory, work seamlessly with any DPLL-based solver). Like Thong et al., we forgo clause lookup and allow clauses to share variables within the same partition. However, we still achieve Davis et al.'s high degree of parallelism by placing clauses directly in clause processors (explained in Section \ref{sec:accelerator}).
\par SAT instances larger than the available number of Clause Processors (CPs) are partitioned, stored in external memory (i.e., software) and hot-swapped into the BCP coprocessors as required during runtime. Solvable instance size is limited only by the GPP's RAM, not on-chip FPGA memory. We deploy our solution on the Zynq chip, and available here\footnote{\url{https://github.com/harigovind1998/FPGA_BCP_acceleration}} for use. To our knowledge, this is the first open-source hardware-accelerated SAT solver.

\subsection{The BCP accelerator architecture} \label{sec:accelerator}
Figure \ref{fig:processor_fpga_interface} illustrates our approach, comprising a GPP and an FPGA accelerated BCP coprocessor. The GPP executes DPLL's remaining elements (decisions, backtrack, etc.), partitions large SAT instances (explained in Section \ref{sec:partition}) and swaps partitions into hardware as required. Its default state is idle, awaiting instructions to execute. Once a decision is received, the systems loops until all unit clauses are exhausted.  The BCP coprocessor, depicted in Figure \ref{fig:processor_fpga_interface}, comprises a control unit (1), an array of clause processors (2) and an implication selector (3). The central control unit communicates directly with the GPP and each CP. Based on the received GPP command, it loads clauses into CPs, broadcasts decisions, or clears assignments during backtrack. At its core, the BCP coprocessor consists of an array of CPs that calculate decision and implication results in parallel. CPs store clauses as an array of literals maintain a local copy of each literal's respective variable assignment. Partitions are hot-swapped into FPGA by overwriting a CPs array of literals with the literals of the new clause. Variable assignments are updated during decisions and BCP, and cleared during backtrack. Finally, the implication selector chooses a single implication to propagate when multiple implications arise as a result of BCP. Rather than using an explicit implication conflict detector, as done by Davis et al \cite{4555925}, we propagate the chosen implication, and identify conflicts during evaluation.

\subsection{Formulae partitioning} \label{sec:partition}
SAT instances contain an arbitrary number of variables and clauses. The problem size solvable on FPGA is limited by its available Configurable Logic Block (CLB) and memory, and requires large problems be partitioned into smaller manageable sizes. Partitions are stored in the GPP, and swapped into FPGA during run time by overwriting CPs clauses. BCP is performed individually on each partition, and implications are relayed back to the GPP. Implications are subsequently propagated to other partitions. We aim to make partitions as large as possible, limited by the coprocessor's clause and variable threshold. Consider Equation \ref{eqn:formula}, composed of four clauses, and an instance of our coprocessor that supports two clauses and three variables. Equation \ref{eqn:variation_1} and \ref{eqn:variation_2} outline the two possible ways to partition Equation \ref{eqn:formula}. Equation \ref{eqn:variation_1} describes a scenario where the partitions reach the clause limit, while the Equation \ref{eqn:variation_2} reaches the variable limit.

\begin{equation} \label{eqn:formula}
    f = (\neg a \lor b \lor \neg c) \wedge (a \lor \neg b \lor \neg c) \wedge (\neg d \lor e \lor f) \wedge (d \lor e \lor f)
\end{equation}

\begin{multline} \label{eqn:variation_1}
    variation\_1=\{\{(\neg a \lor b \lor \neg c) \wedge (a \lor \neg b \lor \neg c)\}, \\\{(\neg d \lor e \lor f) \wedge (d \lor e \lor f)\}\}
\end{multline}
\begin{multline} \label{eqn:variation_2}
    variation\_2=\{\{(\neg a \lor b \lor \neg c)\},\{(a \lor \neg b \lor \neg c)\},\\\{(\neg d \lor e \lor f)\},\{(d \lor e \lor f)\}\}
\end{multline}

\begin{figure}
    \centering
    \includegraphics[width=0.49\textwidth]{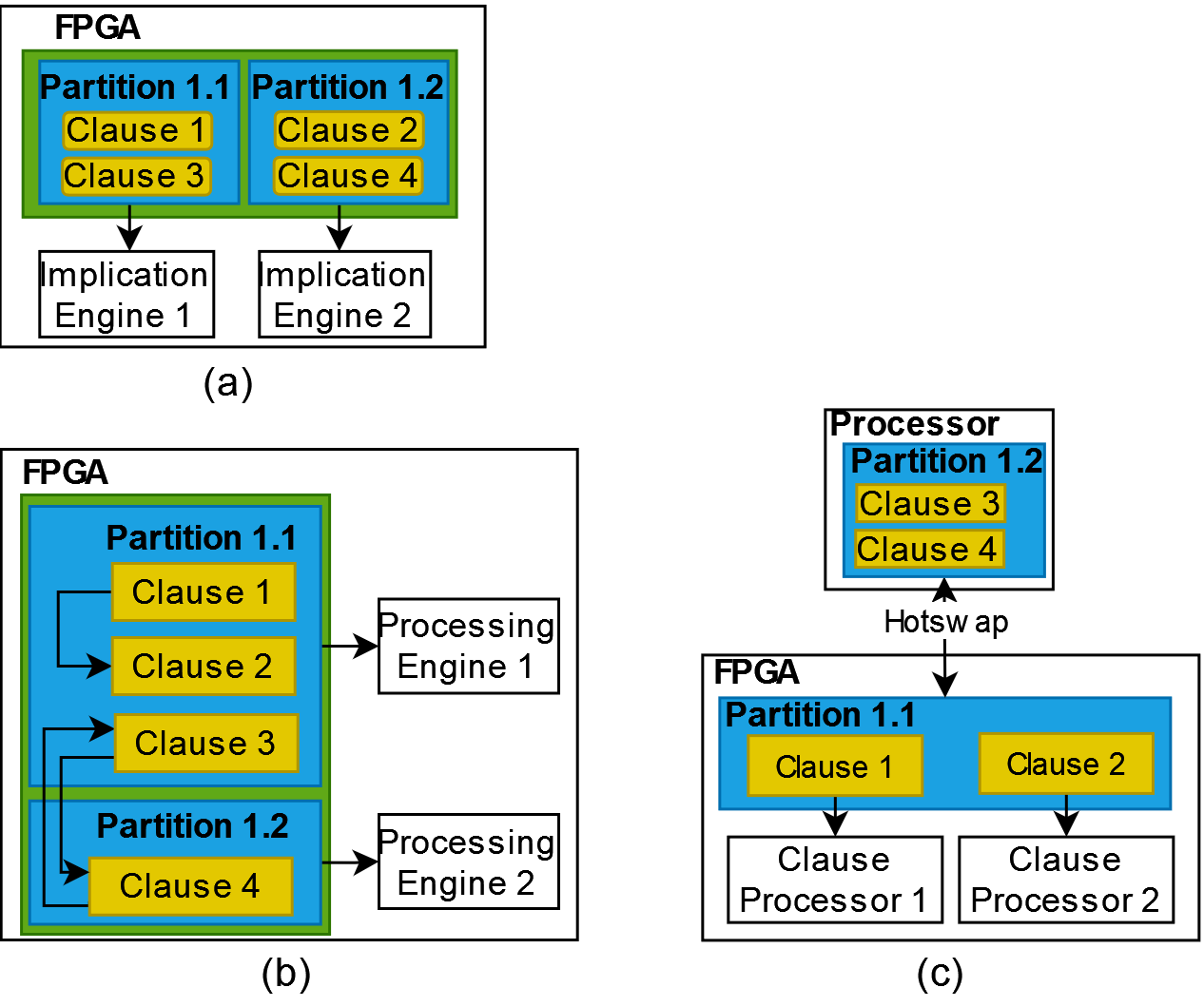}
    \caption{(a) Davis et al. store the formula directly on FPGA. Clauses within partitions contain no shared variables, and partitions are mapped directly to Implication Engines. (b) Thong et al. store formula directly on FPGA. Clauses are linked to other clauses with shared variables and are processed sequentially. (c) Formula stored in external memory ("software" view). Clauses in partitions mapped directly to Clause Processors, and hot-swapped as required.}
    \label{fig:only_arch}
\end{figure}

Results (refer to Section \ref{sec:results}) indicate that partitioning is a bottleneck in our approach. Performance improvement is dictated by the amount of required partition swapping and the number of unused CPs (occurs when the number of clauses in a partition is less than the available number of CPs). Thus, performance improvement is observed with certain partition assignments, while others lead to performance degradation. System performance can be improved by developing a more effective partitioning algorithm, but beyond the scope of this paper and reserved for future work.

\subsection{Execution}
\begin{figure}
\includegraphics[width=0.48\textwidth]{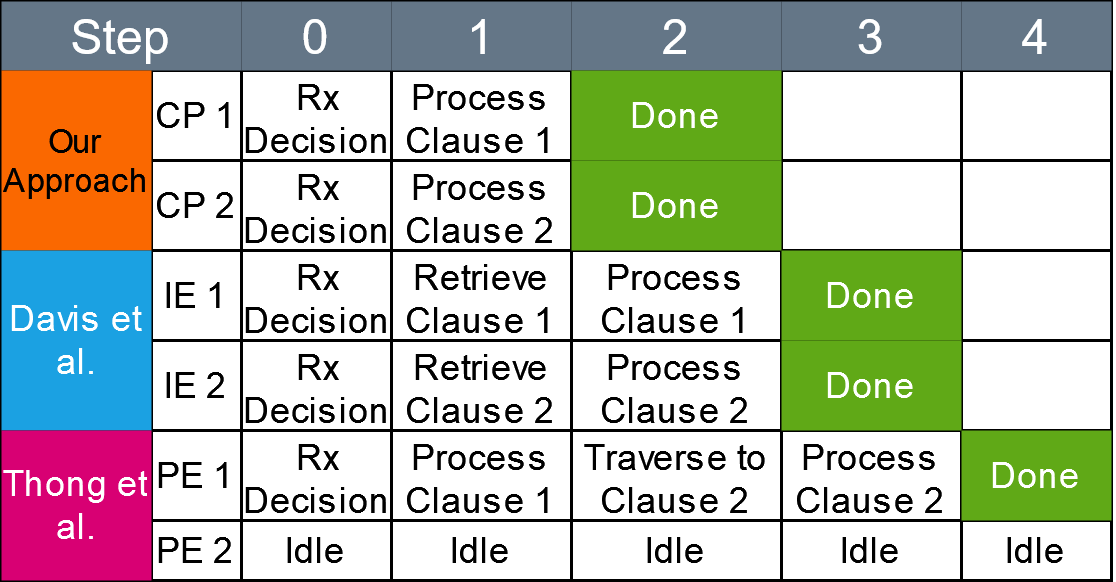}
\centering
\caption{Execution steps of each described approach.}
\label{fig:exec}
\end{figure}

Each clause processor is only associated with a single clause; thus, no clause look-up or traversal is required to retrieve the affected clause for processing. All clauses on the FPGA are processed in parallel as soon as a decision is received. Consider Equation \ref{eqn:formula}'s mapping of partitions to hardware as presented in Figure \ref{fig:only_arch}. Figure \ref{fig:exec} summarizes the execution stages for Davis et al.'s, Thong et al.'s and our approach for the theoretical execution for a decision of variable $a$.  In our approach, clauses 1 and 2 are processed by Clause Processor 1 and 2 in parallel once the decision is received. Since clauses 3 and 4 do not contain variable $a$, Partition 2 remains in external memory and is not processed. Though Davis et al. also process clause 1 and 2 in parallel, each Implication Engine first performs a clause look-up to retrieve the affected clause. Results of the decision on the affected clause are then calculated. Thong et al.'s approach starts BCP on Processing Engine 1. After clause 1 is processed Processing Engine 1 traverses to clause 2. In the manner, clauses in a partition are processed sequentially. Execution concludes after computing Partitions 1's final element, clause 2. Processing Engine 2 remains idle for the entire duration as clauses in partition 2 do not contain variable $a$.

\subsection{Processor-FPGA interface}
The BCP coprocessor implements the Advanced eXtensible Interface 4-Lite (AXI-Lite) IP interface, acting as a subordinate to a processor (AXI master). Using AXI, the processor writes directly to the coprocessor's registers to send instructions and data, and continues polling for status updates and new implication until the coprocessor completes. 
\par Status changes dictate DPLL's flow, either allowing the search to continue assigning additional variables, or triggers backtracking on conflicts. A copy of all the implications are saved on the processor to avoid re-assigning implied variables, and further propagated to the remaining partitions.

\section{Experiments and Results}\label{sec:results}

On a Xilinx Zynq chip with total capacity of 14400 LUTs and 28800 FF, our solution supports 224 parallel Clause Processors and 63 variables. We achieve a clock frequency of 106.66 MHz, utilizing 647 LUTRAM of on-chip memory, 13151 LUTs, and 11059 FFs.

\begin{table}[]
\begin{tabular}{l|lll|}
\cline{2-4}
                                            & \multicolumn{3}{c|}{Millions of BCP/s}                                                                                     \\ \hline
\multicolumn{1}{|l|}{\textbf{SAT Instance}} & \multicolumn{1}{l|}{\textbf{Davis et al    \cite{4555925}}} & \multicolumn{1}{l|}{\textbf{\makecell{Thong et al \\  \cite{7372575}}}} & \textbf{Our Design} \\ \hline
\multicolumn{1}{|l|}{bmc-galileo-8}         & \multicolumn{1}{l|}{40}                    & \multicolumn{1}{l|}{102}                                & 175              \\ \hline
    \multicolumn{1}{|l|}{bmc-ibm-12}            & \multicolumn{1}{l|}{33}                    & \multicolumn{1}{l|}{150}                                & 169               \\ \hline
\end{tabular}
\caption{Comparison of BCP engine throughput (BCPs/s) with related work. Results reflect maximum theoretical throughput, achieved only data is fully available to BCP engines.}
\label{tab:SOTA}
\end{table}

\begin{figure}
\includegraphics[width=0.48\textwidth]{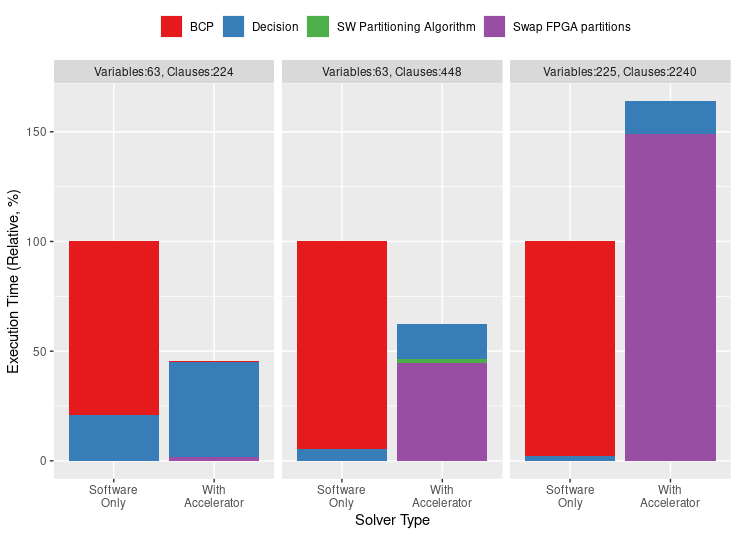}
\centering
\caption{Breakdown of the total execution time across constituent components.}
\label{fig:ExecBreakDown}
\end{figure}

\par
\begin{figure}[t!]
\includegraphics[width=0.48\textwidth]{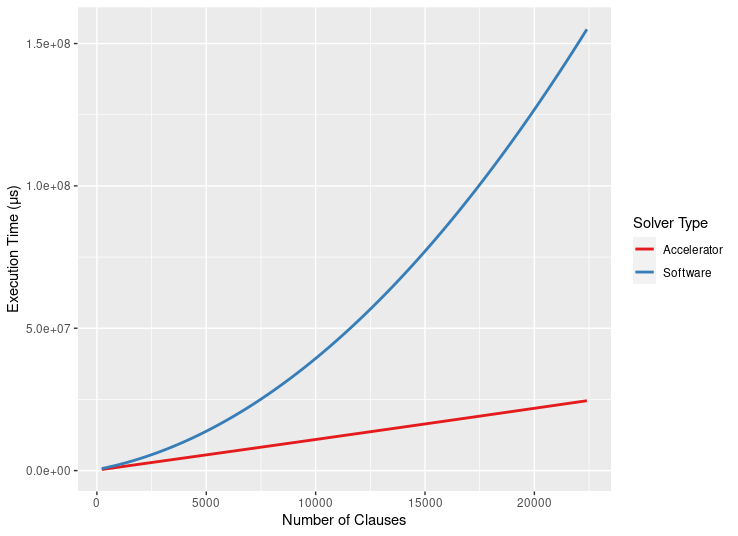}
\centering
\caption{Effect of increasing clauses size on total execution time, for 63 variables.}
\label{fig:BCPvsNumOfClauses}
\end{figure}

\begin{figure}[t!]
\includegraphics[width=0.48\textwidth]{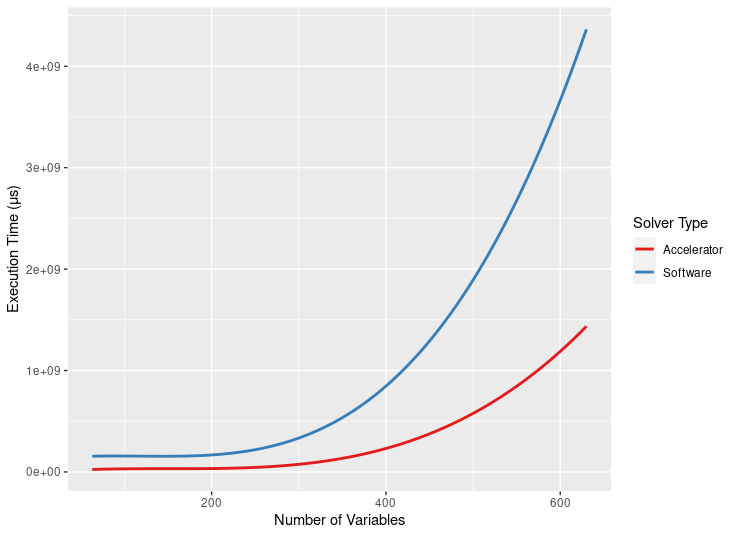}
\centering
\caption{Effect of increasing variables size on total execution time, for 22400 clauses.}
\label{fig:BCPvsVariableNumber}
\end{figure}

 Related work calculates throughput (in BPCs performed per second), assuming full data availability: i.e., not taking into account software execution and communication/data transfer latency. Whilst this is a useful metric to assess hardware performance in isolation (and we report equivalent results in Table \ref{tab:SOTA}), it does not accurately depict system performance; to do so, we break down the full execution in Figure \ref{fig:ExecBreakDown} and evaluate speedup over vanilla software implementation, evaluating combinations of clause and variable sizes, with speedup depicted in Table \ref{tab:matrix} for meaningful combinations. For each combination, we also depict real throughput, in the form of BCPs/s averaged over total execution time (63 variables and 224 clauses is the theoretical upper bound, without the need for hot swapping). To evaluate the different effects of clause/variable sizes on execution, we fix one and vary the other, measuring total execution time: results are depicted in Figures \ref{fig:BCPvsNumOfClauses} and \ref{fig:BCPvsVariableNumber}.

\begin{table}[]
\setlength\tabcolsep{1.5pt}
\centering
\begin{tabular}{cl|llll|}
\cline{3-6}
\multicolumn{1}{l}{}  &   & \multicolumn{4}{c|}{\textbf{Variables}}\\
\cline{3-6} 
\multicolumn{1}{l}{} &  \textbf{}   & \multicolumn{1}{l|}{\textbf{63}} & \multicolumn{1}{l|}{\textbf{126}} & \multicolumn{1}{l|}{\textbf{252}}                     & \textbf{630} \\ 
\hline
\multicolumn{1}{|c|}{\multirow{4}{*}{\rotatebox[origin=c]{90}{\textbf{Clauses}}}}  & \textbf{224}                                                           & \multicolumn{1}{l|}{\makecell{362M BCP/s\\ 2.2x}}     & \multicolumn{1}{l|}{\makecell{17K BCP/s\\0.17x}}   & \multicolumn{1}{l|}{NA}                         & NA\\ \cline{2-6} 
\multicolumn{1}{|c|}{}                                            & \textbf{448}   & \multicolumn{1}{l|}{\makecell{702K BCP/s\\1.6x}}                    & \multicolumn{1}{l|}{\makecell{21K BCP/s\\0.21x}}   & \multicolumn{1}{l|}{\makecell{13K BCP/s\\0.08x}}   & NA\\ \cline{2-6} 
\multicolumn{1}{|c|}{}                                            & \textbf{2240}  & \multicolumn{1}{l|}{\makecell{441K BCP/s\\1.91x}}                   & \multicolumn{1}{l|}{\makecell{22K BCP/s\\1.26x}}   & \multicolumn{1}{l|}{\makecell{16K BCP/s\\0.61x}}   & \makecell{12K BCP/s\\0.10x} \\ \cline{2-6} 
\multicolumn{1}{|c|}{}                                            & \textbf{22400} & \multicolumn{1}{l|}{\makecell{313K BCP/s\\6.32x}}                   & \multicolumn{1}{l|}{\makecell{20K BCP/s\\5.04x}}   & \multicolumn{1}{l|}{\makecell{16K BCP/s\\4.86x}}   & \makecell{14K BCP/s\\3.31x}  \\ \hline
\end{tabular}
\caption{Varied clause/variable sizes and their impact on the relative speedup of hardware/software and the effective throughput of BCP engines.}
\label{tab:matrix}
\end{table}

\section{Conclusions}\label{sec:conclusions}

We described a SAT-solver hardware-accelerated architecture that outperforms state of the art by hot-swapping clause assignment at runtime, making efficient use of FPGA resources. Our solution prototype, on a Xilinx Zynq chip, is available in open-source. Practitioners may use the presented solution in their designs, whenever a problem is encoded in SAT form and performance is critical.
\par An important open question remains: our performance is constrained by how clauses are partitioned. A partitioning scheme that minimizes the distribution of variables among clauses will minimize runtime swapping, resulting in improved execution. However, how to best partition a formula to achieve this is not yet known. Future work must formulate this challenge as an optimization problem, and methods for its efficient solution must be devised. Once that is achieved, they can be applied (offline) prior to deployment on our architecture.

\begin{acks}
We acknowledge the support of the Natural Sciences and Engineering Research Council of Canada (NSERC).
\end{acks}

\bibliographystyle{ACM-Reference-Format}
\bibliography{refs}

\end{document}